\documentclass[final]{svjour3}
\usepackage{graphicx}
\usepackage{rotating}
\usepackage{amssymb}
\usepackage{mathptmx}
\usepackage[numbers]{natbib}
\usepackage{mathtools}  
\usepackage{xfrac}  
\usepackage{xcolor}

\makeatletter
\journalname{Journal of Low Temperature Physics}

\bibpunct{}{}{,}{s}{}{,}

\begin{document}

\newcommand{\tcl}{\ensuremath{T_{CL}}}
\newcommand{\popt}{\ensuremath{P_{opt}(\tcl)}}
\newcommand{\psat}{\ensuremath{P_{sat}}}
\newcommand{\psatd}{\ensuremath{\psat^{\,d}}}
\newcommand{\tbath}{\ensuremath{T_{bath}}}
\newcommand{\pninety}{\ensuremath{P_{90}}}
\newcommand{\pninetyd}{\ensuremath{\pninety^{\,d}}}
\newcommand{\hdblarrow}{H\makebox[0.9ex][l]{$\downdownarrows$}-}

\title{Simons Observatory Focal-Plane Module: In-lab Testing and Characterization Program}

\author{Yuhan Wang$^{1}$
    $\cdot\;$ Kaiwen Zheng$^1$
    \and Zachary Atkins$^1$
    \and Jason Austermann$^2$
    \and Tanay Bhandarkar$^3$
    \and Steve K. Choi$^{4,5}$
    \and Shannon M. Duff$^2$
    \and Daniel Dutcher$^1$
    \and Nicholas Galitzki$^6$
    \and Erin Healy$^1$
    \and Zachary B. Huber$^4$
    \and Johannes Hubmayr$^2$
    \and Bradley R. Johnson$^7$
    \and Jack Lashner$^{8}$
    \and Yaqiong Li$^{4,9}$
    \and Heather McCarrick$^1$
    \and Michael D. Niemack$^{4,5,9}$
    \and Joseph Seibert$^{6}$
    \and Maximiliano Silva-Feaver$^{6}$
    \and Rita Sonka$^1$
    \and Suzanne T. Staggs$^1$
    \and Eve Vavagiakis$^4$
    \and Zhilei Xu$^{10}$ }

\institute{1. Joseph Henry Laboratories of Physics, Jadwin Hall, Princeton University, Princeton, NJ 08544, USA \\
2. Quantum Sensors Group, NIST, 325 Broadway, Boulder, CO 80305, USA \\
3. Department of Physics and Astronomy, University of Pennsylvania, 209 S 33rd St. Philadelphia, PA 19104, USA\\
4. Department of Physics, Cornell University, Ithaca, NY 14853, USA\\
5. Department of Astronomy, Cornell University, Ithaca, NY 14853, USA\\
6. Department of Physics, University of California San Diego, La Jolla, CA 92093 USA\\
7. Department of Astronomy, University of Virginia, Charlottesville, VA 22904, USA \\
8. Department of Physics and Astronomy, University of Southern California, Los Angeles, CA 90089, USA\\
9. Kavli Institute at Cornell for Nanoscale Science, Cornell University, Ithaca, NY 14853, USA\\
10. MIT Kavli Institute, Massachusetts Institute of Technology, 77 Massachusetts Avenue, Cambridge, MA 02139, USA\\
\email{yuhanw@princeton.edu, kaiwenz@princeton.edu}}

\maketitle

\begin{abstract}

The Simons Observatory (SO) is a ground-based cosmic microwave background instrument to be sited in the Atacama Desert in Chile.  SO will deploy 60,000 transition-edge sensor (TES) bolometers in 49 separate focal-plane modules across a suite of four telescopes covering three dichroic bands termed low frequency (LF), mid frequency (MF) and ultra-high frequency (UHF). Each MF and UHF focal-plane module packages 1720 feedhorn-coupled detectors with cryogenic components for highly multiplexed readout using microwave SQUID multiplexing.
In this paper we describe the testing program we have developed for high-throughput validation of modules after they are assembled. The validation requires measurements of the yield, saturation powers, time constants, noise properties and optical efficiencies. Additional measurements will be performed for further characterizations as needed. We describe the methods developed and demonstrate preliminary results from the initial testing of a prototype module. 


\end{abstract}

\section{Introduction}

The Simons Observatory (SO) is a suite of ground-based telescopes to be sited in the Atacama Desert in Chile focusing on measuring the temperature and polarization anisotropy of the cosmic microwave background (CMB). The initial deployment of SO will consist of one 6 m large aperture telescope (LAT) and three 0.5 m small aperture telescopes (SATs). Over 60,000 transition-edge sensor (TES) bolometers will be deployed to achieve the target mapping speed\cite{forcast}. The TES bolometers and readout circuitry based on microwave SQUID multiplexers\cite{dober2021} will be packaged into 49 separate universal focal-plane modules (UFMs) spanning six frequency bands from 30 GHz to 280 GHz. The 30$/$40 GHz low frequency (LF) UFMs use lenslet-coupled sinuous antennas, while the 90$/$150 GHz mid frequency (MF) and the 220$/$280 GHz ultra-high frequency (UHF) UFMs implement horn-coupled orthomode transducers\cite{{so20},{Galitzki2018}}. Each MF and UHF focal-plane module packages 1720 optical detectors and 36 dark bolometers. Further details of the SO focal-plane modules can be found in McCarrick, {\it et al}. (2021). \cite{mccarrick2021}

During observations, UFMs will be mounted to the 100 mK stages of the LAT and SAT receivers. Each UFM is connected to two cold readout chains, reading out 1820 readout channels in a 2x910 multiplexing configuration. Each readout chain consists of cryogenic coaxial cables and various RF components selected with considerations of system linearity,  noise performance, and  thermal power dissipation \cite{rao20}. The SLAC Superconducting Microresonator RF (SMuRF) electronics serve as the room temperature readout electronics\cite{henderson18}.

All 49 UFMs deployed in the observatory will first be 
tested and characterized in the laboratory. Our testing program is based on testing runs, each of which yields both dark and optical properties for three UFMs at the same time, in one dilution refrigerator cryostat (DR) using readout chains which are analogous to those to be used in the SO receivers. Section 2 describes the tools and the high-throughput validation methods developed for this pre-deployment testing, and Section 3 describes preliminary testing results.

\section{Methods}
\label{methods}
\subsection{Testing Hardware}
\begin{figure}
\begin{center}
\includegraphics[width=0.9\linewidth, keepaspectratio]{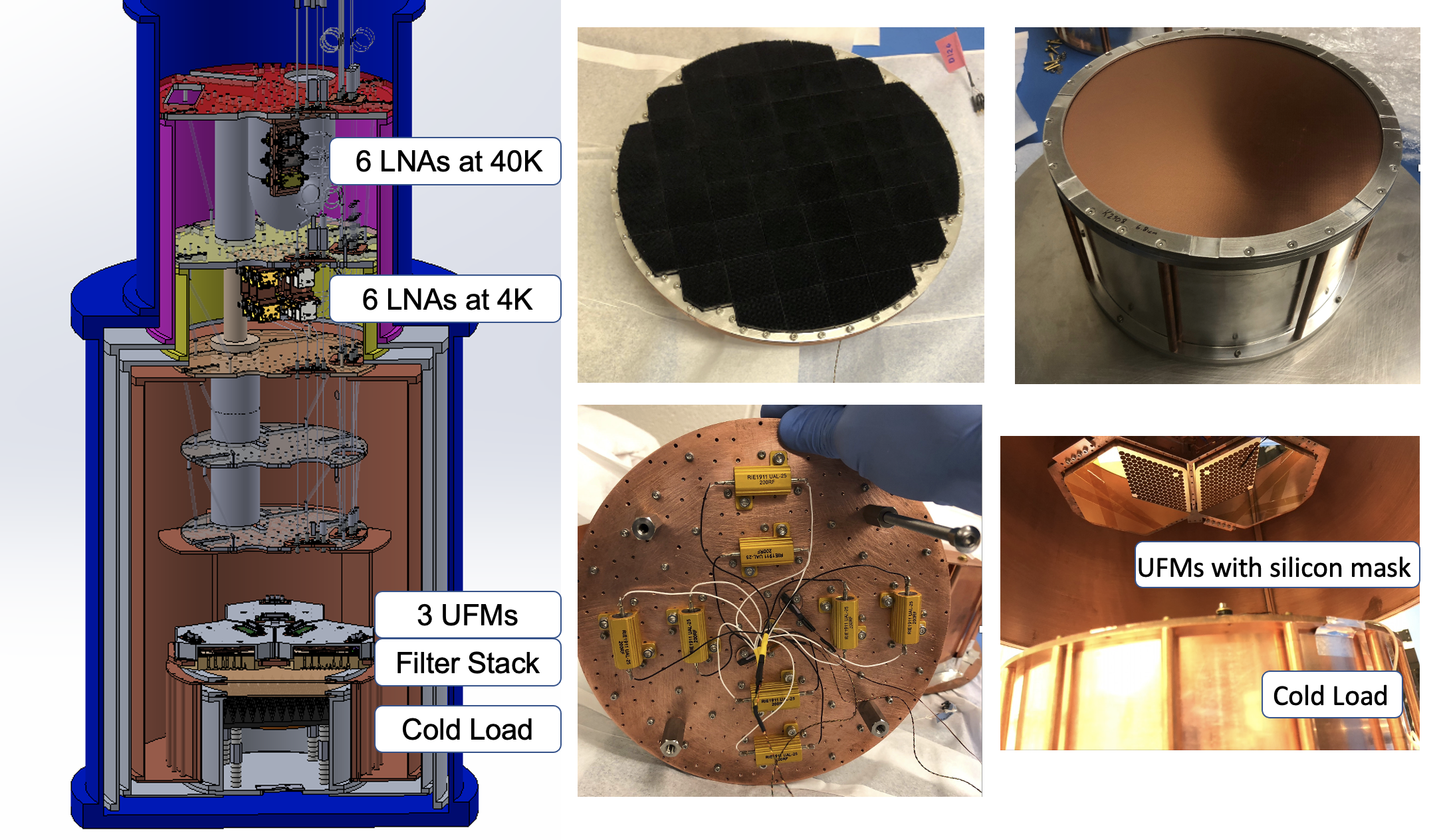}
\caption{{\it Left}: Schematics of the testing setup inside the DR showing:  six readout chains with low-noise amplifiers (LNAs) located at the 40 K and 4 K stages;  three UFMs mounted under the DR mixing chamber;  the free space filter stack;  and the cold load. 
{\it Top middle}: Photograph of one of the DR cold loads, featuring  metamaterial absorbing tiles\cite{xu21}.
{\it Bottom middle}: Heaters and thermometers beneath the cold load plate for temperature control. 
{\it Top right}: The mechanical housing and metal-mesh low pass edge (LPE) filters for the cold load.
{\it Bottom right}: UFMs mounted inside the DR facing the internal cold load, showing the masks that cover two-thirds of the detectors so that both dark and optical properties can be sampled for each UFM.
(Color figure online.)}
\label{testing_setup}
\end{center}
\end{figure}

The laboratory characterization and testing of the UFMs takes place in two Oxford Instruments Triton 200 DRs at Princeton University and one Bluefors LD DR at Cornell University.\footnote{The Bluefors DR has small differences from the others;  in what follows,  we focus on the Oxford systems for concreteness. } Each fridge can test three UFMs in the same cool-down as shown in the left panel of Figure~\ref{testing_setup}. The UFMs are mounted on a copper rack, which provides similar thermal contact area as the focal-plane mounting plates for the UFMs in the LAT and SAT receivers.
The copper rack is bolted and heat sunk to the bottom of the DR mixing chamber (MXC), reaching a temperature that agrees with the MXC temperature to within 1 mK. Therefore, we use thermal control of the MXC, provided by an AC372 Lakeshore controller, to enable UFM characterizations at different bath temperatures \tbath, which we define with a thermometer permanently mounted on the MXC of each DR. We can ramp \tbath\ from 60~mK to 250~mK, covering both the planned UFM operating temperature of 100 mK and the targeted device critical temperature $T_c$ of 160 mK.

An internal cold load is designed and fabricated to act as a cryogenic thermal source. The cold load structure is shown in the middle panel of Figure~\ref{testing_setup}. The cold load is made of pyramids of emissive metamaterial microwave absorber\cite{xu21} bolted to a 20~cm diameter aluminum disk. Heaters and diode thermometers are mounted behind the cold load plate. During in-lab testing, four thermally-tuned standoffs support the cold load above the bottom of the 4 K shield of a DR as shown in Figure~\ref{testing_setup}. The standoffs define the thermal conductance of the cold load to the DR 4 K stage and have been carefully tuned for sufficient isolation. An LS336 Lakeshore controller records and commands changes in the cold load temperature from 8 K to 24 K.

Two metal-mesh low-pass edge (LPE) filters\cite{ade2006} are mounted along the optical path from the cold load to the feedhorn apertures of the UFMs, one at the 4K stage and the other at the 1 K stage, as shown in Figure~\ref{testing_setup}. The LPE filters help define the highest-frequency band edge for the dichroic UFMs and reduce sensitivity to possible blue leaks. We use separate LPE filter stacks for each of the frequency ranges: LF, MF and UHF.

The plan is that all UFMs will be tested in lab with the internal cold load before being deployed. Copper or gold-plated silicon masks are attached over the feedhorn apertures as shown in the bottom right of Figure~\ref{testing_setup}. Each mask covers two thirds of the detectors in a UFM. We refer to these masked detectors as ``dark" detectors and measure their properties as described in Section 2.2. The mask exposes one third of the detectors to the optical loading from the cold load, which we use to estimate their optical efficiencies in Section 2.3. The measurements of dark detectors are also used to decouple the optical loading and thermal coupling in optical efficiency calculations. Although occasional tests have been performed without the internal cold load, the standard validation plan does not include such dark runs.

The UFMs are read out by the same type of RF chains and warm electronics as will be used in the field.
A series of attenuators on the input side of the RF chain helps achieve the optimal input power for the resonators. The output side of the cold readout chain uses two-stage amplification with one low-noise RF amplifier (LNA) at the 40 K stage and another at the 4 K stage to optimize the linearity and maintain low noise. The SMuRF electronics provide the LNA and detector biases, and control the operation of all SQUID coupled resonators.

\subsection{Characterization of electrical and thermal properties}
\label{sec:electricalthermal}
As described above, the dark (masked) detectors are used to validate the TES electrothermal properties\cite{irwin05} of each UFM. The properties we validate in lab are saturation powers (section~\ref{sec:thermal}), biasability (section~\ref{sec:biasability}), time constants (section ~\ref{sec:biasstep}), and noise performance (section~\ref{sec:noise}). We also measure the detector thermal conductances to the thermal bath and the detector responsivities since they affect the thermal noise and the sensitivity of the array. The TES critical temperature $T_{c}$ and the normal resistance ${R_n}$ are also measured to trace fabrication stability and uniformity.

\subsubsection{Thermal properties}
\label{sec:thermal}

Each TES resistor (with $R_n \sim 8~\mbox{m}\Omega$) is in parallel with a shunt resistor (with $R_s \sim 400~\mu\Omega$) so that it can be voltage-biased in its transition by a current into the circuit.  That current is proportional to $V_{bias}$ provided by the SMuRF electronics.  We measure TES I-V curves by sweeping $V_{bias}$ from high to low while reading the current in the TES.
The TES resistances can be calculated at each ${V_{bias}}$ and the TES normal resistance ${R_n}$ can be extracted by linear fitting of the normal part of the corresponding I-V curve.

We define saturation power
\pninety\ as the electrical power through a TES at which its operating resistance equals $90{\%}$ of ${R_n}$ as determined from an I-V curve.  We use a superscript $d$ when the measurement is made on a dark detector. I-V curves taken across a range of bath temperatures are used to estimate $T_{c}$ and the thermal conductance $G$ using the method outlined by Sudiwala, Griffin and Woodcraft (2002).\cite{sudiwala2002,crowley16,choi18}  For the dark detectors, we fit 
\begin{equation}
   \pninetyd(\tbath) = \kappa(T_c^{n}-T_{bath}^{n}), \label{eq:1}
\end{equation}
extract the parameters $T_c$, $\kappa$ and $n$, and estimate $G=n\kappa T_c^{n-1}$.  We also  track \pninetyd\ at $T_{bath} = 100$~mK. 

\subsubsection{Biasability}
\label{sec:biasability}

Each UFM has 12 detector bias lines, each providing voltage biasing to $\sim 150$ detectors. During observations, variations in the sky conditions result in varying optical load on the detectors. Whether detectors located on the same bias line can be biased into the transition simultaneously under various loading conditions impacts the overall detector yield. 
We quantify what we call the UFM "biasability" by the percentage of TESes with resistances between 30 and 70$\%$ of ${R_n}$ at a single (optimal) bias voltage on each of the 12 bias lines, and measure it from I-V curves.

\subsubsection{Time constants and responsivity}
\label{sec:biasstep}
The time constants and detector responsivity are measured in lab using bias step measurements\cite{niemack08}, for which a small-amplitude square wave is added to the DC bias level on each detector bias line. The step function can be thought of as a two-point I-V curve. The amplitude of the resulting change in TES current can be used to calculate the TES responsivity. In addition, since the detector time constant functions as an effective low pass filter on the small-amplitude square wave,  the time constant can be extracted by fitting the exponential settling of the TES response in time domain. A comparison of bias step and complex impedance measurements of the effective time constant can be found in Cothard, {\it et al}. (2020). \cite{cothard20} 

\subsubsection{Noise in transition}
\label{sec:noise}

A key indicator of the UFM quality is its noise performance. To characterize the detector noise, we first apply sufficient bias voltage to drive the detectors normal, and then step down in voltage, collecting two-minute time streams for each detector at every bias voltage step. The white noise level for each detector is approximated using the median of the amplitude spectral density of the time stream data between 5 and 50 Hz. This frequency range is selected to avoid both the low frequency region which can be contaminated by thermal $1/f$ in the lab and the high-frequency roll-off defined by the anti-aliasing filter used in the SMuRF electronics.

The measured noise levels with TESes in the normal, superconducting and in-transition states for dark detectors and detectors exposed to the cold load are then compared to the white noise levels 
expected for those six cases. The array sensitivity can be 
estimated from the measured TES in-transition noise in combination with the responsivity measurements described in section \ref{sec:biasstep}.

\subsection{Characterization of optical properties}

The optical property we validate in lab is the optical efficiency, defined as the fraction of the incident power which is absorbed and measured by a detector. For each detector location, the incident power \popt\ at each temperature \tcl\ of the cold load
is calculated by integrating its blackbody emission over the beam and the expected detector bandpass including the LPE filters.\cite{{crowley16},{choi18}}

We step \tcl\ from $\sim 8$K to $\sim 20$K with the bath temperature fixed, typically at 100~mK. At each \tcl, we measure \pninety\ from I-V curves for the unmasked detectors. We find the optical efficiency $\eta$ as the absolute value of the slope when  fitting \pninety\  versus \popt\ to a line: 
\begin{equation}\label{eq:4}
\pninety(\tcl)  = \pninetyd -\eta\popt.
\end{equation}

Note that the offset of the line is an estimate of the saturation power in the absence of optical loading (as in equation~\ref{eq:1}).

The cold load can also cause thermal loading on the UFM that increases with \tcl. If left uncorrected, it would result in an overestimation of the optical efficiency. This  effect is corrected by subtracting the median \pninetyd\ for the dark detectors from
\pninety\ for the optical detectors at each value of \tcl\ before performing the line fit.

\section{Results and Discussion}

\begin{figure}
\begin{center}
\includegraphics[width=1\linewidth, keepaspectratio]{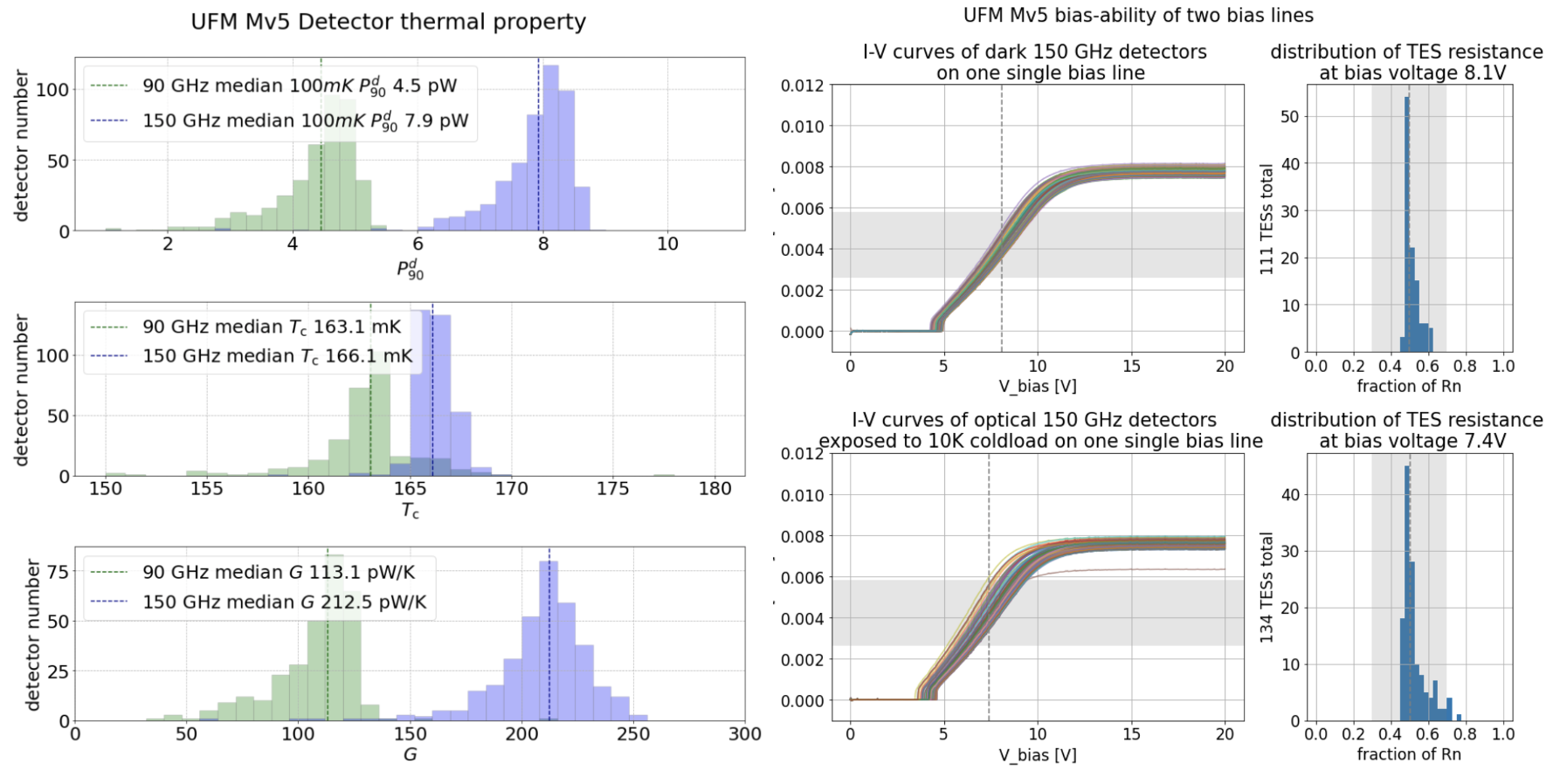}
\caption{Thermal and electrical properties of the prototype UFM Mv5 demonstrating testing methods and capabilities.
{\it Left}: Saturation powers at 100mK, critical temperatures and thermal conductances for the dark detectors of the prototype UFM-Mv5.
{\it Middle and Right}: Biasability of UFM-Mv5, showing two typical bias lines of twelve. The targeted region of 30 to $70{\%}$ ${R_n}$ is indicated with grey shading. The detectors represented in the bottom row were exposed to the  
cold load at a temperature of 10K. For each bias line, a range of possible $V_{bias}$ values could be used to put most detectors into the target zone of resistance.
\label{fig:thermal_properties.png}
(Color figure online.)}
\end{center}
\end{figure}

To demonstrate the in-lab testing program presented here, we present some preliminary results from testing a prototype MF UFM designated UFM-Mv5. Detailed results on laboratory characterization of SO UFMs for deployment will be presented in future work.

The thermal properties for UFM-Mv5 are shown on the left side of Figure~\ref{fig:thermal_properties.png}. The measured $T_{c}$, \pninetyd\ at 100 mK, 
and $G$ for the dark detectors in UFM-Mv5 
can be found in Table \ref{table:1}, along with 
the noise equivalent power (NEP) for dark detectors biased at $50{\%}$ ${R_n}$,  measured at 100 mK.

\begin{table}[h!]
\centering
\begin{tabular}{ |p{1.5cm}||p{1.5cm}|p{2.5cm}|p{2cm}|p{2cm}| }
 \hline
 \multicolumn{5}{|c|}{UFM Mv5} \\
 \hline
detector frequency (GHz) & $T_{c}$ (mK) & \pninetyd\  (pW) at 100 mK& $G$ (pW/K)& $NEP$ (aW/rtHz) at 100 mK\\
 \hline
 90   &   163.1 $\pm$ 2.6   & 4.5 $\pm$ 0.7 & 113.1 $\pm$ 19.2  &    13 $\pm$ 3\\
 \hline
150   &   166.1 $\pm$ 1.0   & 7.9 $\pm$ 0.7 & 212.5 $\pm$ 22.2  &   18 $\pm$ 4\\

 \hline
\end{tabular}
\caption{The medians and standard deviations of thermal parameters measured for each frequency band of the prototype UFM-Mv5 using methods described in section \ref{sec:electricalthermal}. The rightmost column gives the noise equivalent power measurements of dark detectors biased at $50{\%}$ ${R_n}$ at 100 mK.}
\label{table:1}
\end{table}

Examples of the biasability of dark and optical 150 GHz detectors from two different bias lines in UFM-Mv5 are shown on the right side of Figure~\ref{fig:thermal_properties.png}. The plot shows that bias voltages can be chosen such that almost all detectors from the two example bias lines have operating resistances between 30 and $70{\%}$ ${R_n}$.

\begin{figure}
\begin{center}
\includegraphics[width=0.9\linewidth, keepaspectratio]{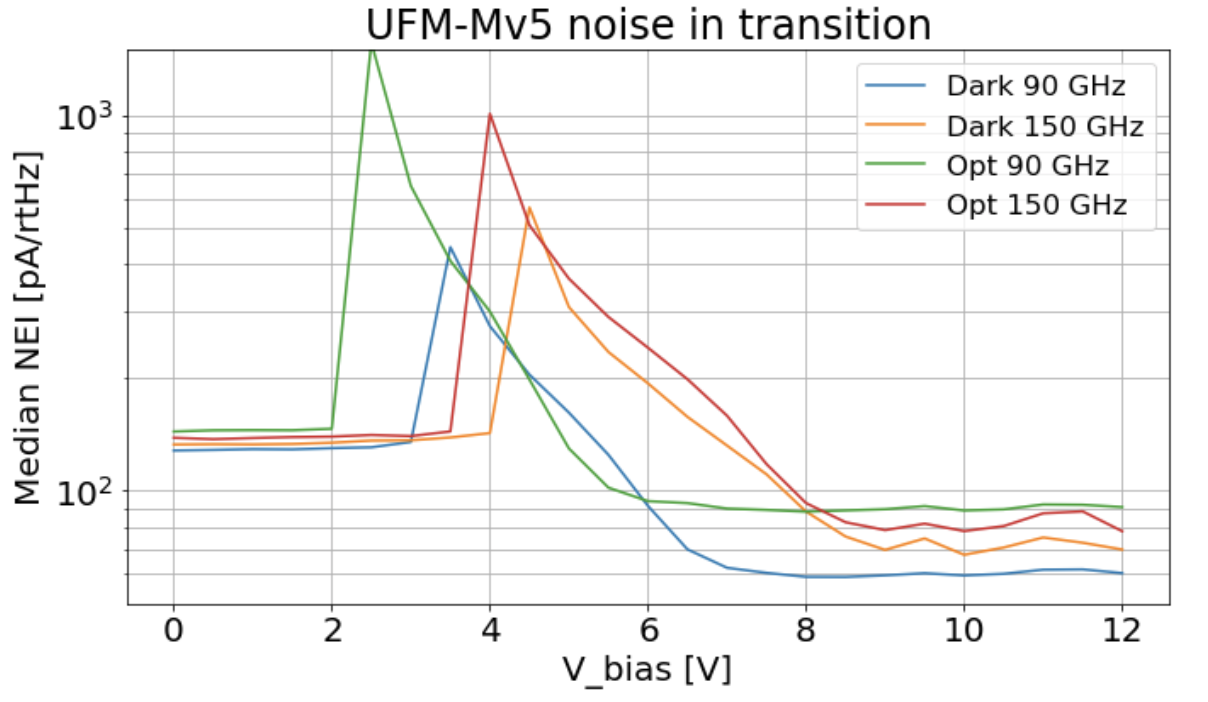}
\caption{Noise equivalent current at different points along the TES transition for dark (masked) detectors and optical detectors (exposed to the 10 K cold load). The plot shows the median measured noise at a range of bias voltages. With higher Johnson noise, the median TES noise in the superconducting state is higher than the noise in the normal state as expected. The noise from photon loading from the internal cold load is apparent when comparing optical and dark detectors' median noises at the same percentage ${R_n}$.
\label{fig:noise.png}
(Color figure online.)}
\end{center}
\end{figure}

Figure~\ref{fig:noise.png} shows noise performance.  
For the current readout system, the Johnson noise in the TES-plus-shunt-resistor loop is inversely proportional to the resistance in the loop.  At low bias voltages, the TES is  superconducting, so the shunt resistance dominates, and the Johnson noise level is higher than when the TES is normal. When the detectors are in transition, noise due to photon loading additionally contributes to the expected TES noise. As can be seen from the right side of Figure~\ref{fig:thermal_properties.png}, to reach the same TES resistance, different bias voltages are needed for dark detectors compared to unmasked detectors. Figure~\ref{fig:noise.png} shows different TES noise levels at a range of bias voltages and confirms the expected noise behaviors discussed above, including a higher noise at the superconducting state than the normal state, and the effect of the photon noise when TESes are in transitions.

\section{Conclusion}
We discussed the in-lab testing and characterization program developed for the focal-plane modules for SO. We described the methods developed to allow simultaneous testing of dark and optical properties of three UFMs at the same frequency in one dilution fridge using six readout chains. We have used the testing program to validate prototype MF UFMs and reported here some preliminary results. We expect a total of 49 arrays to be tested and validated in the same manner before their assembly into the LAT and the SATs. Detailed results from the full testing program, including measurements of optical efficiency, yield and time constants, will be reported in future work.\\
\newline
\newline
\newline

\noindent \small \textbf{Data Availability} The data that support the findings of this study are available from the corresponding authors, YW and KZ, upon reasonable request.

\begin{acknowledgements}
This work was supported in part by a grant from the Simons Foundation (Award 457687, B.K.) and private funding from universities.  
SKC acknowledges support from NSF award AST-2001866. 
YL is supported by KIC Postdoctoral Fellowship.
ZBH is supported by a NASA Space Technology Graduate Research Opportunities Award.
ZX is supported by the Gordon and Betty Moore Foundation through grant GBMF5215 to the Massachusetts Institute of Technology.
\end{acknowledgements}


\begin{thebibliography}{99}

\bibitem{forcast}
The Simons Observatory Collaboration. {\it JCAP}, {02(2019)} 056
DOI: 10.1088/1475-7516/2019/02/056

\bibitem{dober2021}
B. Dober, et al. {\it Appl. Phys. Lett.} \textbf{118}, 062601 (2021), DOI: 10.1063/5.003341

\bibitem{so20}
The Simons Observatory Collaboration. {\it BAAS} \textbf{51}, 147 (2019), https://arxiv.org/abs/ 1907.08284

\bibitem{Galitzki2018}
N. Galitzki, et al. {\it SPIE Proceedings } \textbf{10708}, 1070804 (2018),DOI: 10.1117/ 12.2312985

\bibitem{mccarrick2021}
H. McCarrick and E. Healy, et al. {\it ApJ} \textbf(922) 38 (2021), DOI: 10.3847/1538-4357/ac2232

\bibitem{rao20}
M. S. Rao, M. Silva-Feaver, et al. {\it J. Low Temp. Phys.} \textbf{199}, 807–816 (2020), DOI: 10.1007/ s10909-020-02429-y

\bibitem{henderson18}
S. W. Henderson, et al. {\it SPIE Proceedings } \textbf{10708}, 1070819 (2018), DOI: 10.1117/ 12.2314435

\bibitem{xu21}
Z. Xu and G. E. Chesmore, et al. {\it Appl. Opt.} \textbf{60(4)}, 864-874(2021), DOI: 10.1364/ AO.411711

\bibitem{ade2006}
P. A. R. Ade, G. Pisano, C. Tucker, S. Weaver. {\it SPIE Proceedings }\textbf{6275}, 62750U (2006), DOI: 10.1117/ 12.673162

\bibitem{irwin05}
K. D. Irwin, G. C. Hilton. {\it Topics in Applied Physics} \textbf{99}, 63 (2005), DOI: 10.1007/ 10933596\textunderscore3


\bibitem{sudiwala2002}
R. V. Sudiwala, M. J. Griffin, \& A.L.  Woodcraft. {\it J. Infrared Millim. Terahertz Waves}, \textbf{23}, 545–573 (2002), DOI: 10.1570 / 5.826900

\bibitem{crowley16}
K. T. Crowley, et al. {\it SPIE Proceedings}  \textbf{9914}, 991431 (2016), DOI: 10.1117/ 12.2231999

\bibitem{choi18}
S. K. Choi, et al. {\it J. Low Temp. Phys.} \textbf{193}, 267–275(2018), DOI: 10.1007/ s10909-018-1982-4

\bibitem{niemack08}
M. D. Niemack. Towards Dark Energy:Design, Development, and Preliminary Data from ACT. {\it Ph.D.Thesis} (2008), pp. 110–118

\bibitem{cothard20}
N. F. Cothard, et al. {\it SPIE Proceedings} \textbf{11453}, 11453185 (2020), DOI: 10.1117/ 12.2575912


\end{thebibliography}
\end{document}